\documentclass[%
 reprint,
 amsmath,amssymb,
prb,
]{revtex4-1}

\usepackage{amsmath} 

\usepackage{graphicx}

\usepackage{amsthm}

\newcommand{\be}{\begin{equation}}
\newcommand{\ee}{\end{equation}}
\newcommand{\bea}{\begin{eqnarray}}
\newcommand{\eea}{\end{eqnarray}}

\newcommand{\rme}{{\rm{e}}}

\newcommand{\iGAj}{{j}}

\newcommand{\cliffconj}[1] { \bar{#1} }

 \theoremstyle{definition}
 
 \theoremstyle{remark}

 \numberwithin{equation}{section}

\begin{document}

\preprint{APS/123-QED}

\title{Generalizing the Lorentz transformations}

\author{James M.~Chappell}
\affiliation{School of Electrical and Electronic Engineering, \\ University of Adelaide, SA 5005 \\ Australia}
\email{james.chappell@adelaide.edu.au}

\author{David L.~Berkahn}
\address{School of Electrical and Electronic Engineering, \\ University of Adelaide, SA 5005 \\ Australia}

\author{Nicolangelo Iannella}
\address{School of Mathematical Sciences, \\ University of Nottingham, NG7 2RD, \\ UK}
\address{Institute for Telecommunications Research, \\ The University of South Australia, Mawson Lakes SA 5095, \\ Australia}

\author{John G.~Hartnett}
\address{Institute for Photonics \& Advanced Sensing (IPAS), and the \\ School of Physical Sciences, \\ University of Adelaide, Adelaide SA 5005 \\ Australia}
\email{john.hartnett@adelaide.edu.au}

\author{Azhar Iqbal}
\address{School of Electrical and Electronic Engineering, \\ University of Adelaide, SA 5005 \\ Australia}

\author{Derek Abbott}
\affiliation{School of Electrical and Electronic Engineering, \\ University of Adelaide, SA 5005 \\ Australia}

\date{\today}

\begin{abstract}
In this paper we develop a framework allowing a natural extension of the Lorentz transformations. To begin, we show that by expanding conventional four-dimensional spacetime to eight-dimensions that a natural generalization is indeed obtained.
We then find with these generalized coordinate transformations acting on Maxwell's equations that the electromagnetic field transformations are nevertheless unchanged. We find further, that if we assume the absence of magnetic monopoles, in accordance with Maxwell's theory, our generalized transformations are then restricted to be the conventional ones. While the conventional Lorentz transformations are indeed recovered from our framework, we nevertheless provide a new perspective into why the Lorentz transformations are constrained to be the conventional ones.  Also, this generalized framework may assist in explaining several unresolved questions in electromagnetism as well as to be able to describe quasi magnetic monopoles found in spin-ice systems.
\end{abstract}

\maketitle


\section{Introduction}

It was an unexpected result of the nineteenth century that the Galilean transformations, well established in classical mechanics for observers in relative motion, did not appear to apply to electromagnetic processes.  Indeed, it was shown by Lorentz and Poincar{\'e} that an alternate set of transformations were required to ensure the invariance of Maxwell's equations~\cite{Jackson1998}.  This set of equations, known today as the Lorentz transformations, were then incorporated as a foundational result into Einstein's special theory of relativity~\cite{Jackson1998}.

Various attempts have since been been made to generalize the Lorentz transformations,\cite{Weinberg1980} such as by including transformations to include superluminal velocities\cite{Mignani1973}---however this introduced imaginary quantities with no clear physical interpretation.  Other approaches have included relaxing space isotropy,\cite{Alway1969} which has also been shown to be inconsistent with experiment\cite{Strnad1970} as well as other approaches introducing non-linear transformations.
More exotic suggestions include the introduction of three time dimensions,\cite{cole1982space} although in this case there are difficulties with interpretation. It was concluded by Weinberg, that all these approaches are essentially inapplicable\cite{Weinberg1980}. 

On the other hand, our approach begins with a generalization of spacetime to eight-dimensions through including rotational degrees of freedom into space, while retaining a single time dimension.  We also remain in real space with purely subluminal motion and retain all the usual relativity postulates, including linearity.  This expanded description of spacetime then allows a generalized Lorentz transformation that we are seeking. The transformation retains the invariant interval as well as the form of Maxwell's equations, as required.  We find that with the assumption of the non-existence of monopoles embodied in Maxwell's original equations our generalized transformation is constrained to coincide with the conventional Lorentz group.
Note that other more conventional generalizations are possible such as incorporating provisions for accelerating and rotating frames,\cite{Nelson1987} however we wish to remain in the context of inertial observers.

\subsection{Minkowski spacetime}

To produce the Lorentz transformations, we begin by assuming the conventional four-dimensional spacetime representation, where we have an event 
\be
X = [ c t, \boldsymbol{x} ] ,
\ee
where $ \boldsymbol{x} = x_1 e_1 + x_2 e_2 + x_3 e_3 $ is the space coordinate and $  t $ is the local time in this frame and $ c $ is the canonical speed of light.
Now, Maxwell's equations state that the speed of light $ c = 1/\sqrt{\mu_0 \epsilon_0 } $
is isotropic and the same measured value for all observers.  Hence, for two inertial frames $\mathcal{S} $ and $\mathcal{S}'$ moving with relative velocity $ v $ along the $ x$-axis for example, if a light is flashed at $ t' = t = 0 $ at the coordinate origin, then with an isotropic speed of light, independent of motion,  both observers will find themselves at the center of a spherical light sphere. Hence, we obtain the relation
\be \label{TwoObserverLightCones}
{x'}_1^2 + {x'}_2^2 + {x'}_3^2 - c^2 t^{'2} = x_1^2 + x_2^2 + x_3^2 - c^2 t^2 ,
\ee
between the two observers.
It is well known that the transformations required to satisfy this relation, for an observer moving along the $ x_1 $ axis, are the Lorentz transformations
\bea  \label{spacetimeLorentzTransformations}
x_1' & = & \gamma (x_1 - v t) \\ \nonumber
x_2' & = & x_2 \\ \nonumber
x_3' & = & x_3 \\ \nonumber
t' & = & \gamma (t - v x_1/c^2 ),  \nonumber
\eea
where $ \gamma = \frac{1}{\sqrt{1 - v^2/c^2 }} $.  
Note, that an assumption of the isotropy of the speed of light, in all frames of reference, is required to yield this set of equations. This synchronization scheme is known as the Einstein Simultaneity Convention (ESC), or the Einstein Synchrony Convention. In fact, provided we assume a round trip speed for light of $ c $ during clock synchronization, without requiring any assumptions about the one way speeds, then other simultaneity conventions can be adopted in an equally self consistent manner. The reason for this stems from our apparent inability to measure the one-way speed of light,\cite{weingard1985remark} though we can avoid these issues here by simply adopting the ESC, as did Einstein and hence the name.

\section{Clifford geometric algebra}

An alternate formalism for spacetime is Clifford geometric algebra $ C\ell (\Re^3)$ that provides a generalization of the Gibbs-Heaviside three-vector formalism. In this case we write an eight-dimensional spacetime event 
\be \label{multivector3DSimple}
X = c t + \boldsymbol{x} + \iGAj \boldsymbol{n} + \iGAj c b ,
\ee
where $ \iGAj \boldsymbol{n} = n_1 e_2 e_3 + n_2 e_3 e_1 + n_3 e_1 e_2 $ is a bivector, and $ \iGAj = e_1 e_2 e_3  $ a trivector.  The bivectors have the transformation properties of pseudovectors and thus describe physical quantities such as the magnetic field or angular momentum.  The trivectors are pseudoscalars and so describe helicity or magnetic monopoles. The multivector can thus be viewed as a generalized description of spacetime that now includes the additional components $ \iGAj \boldsymbol{n} + \iGAj c b = \iGAj (\boldsymbol{n} + c b) $ which has the properties of four-spin, conventionally represented by the four-vector $ [ c b , \boldsymbol{n} ] $. Time, defining the radius of the spherically symmetric light sphere $ c t $, is aptly described as a scalar, with the distance $ \boldsymbol{x} $ measured as a fraction of this light distance, being a directed quantity represented as the linear combination of three orthogonal directions.
Note that the eight dimensional Clifford multivector description of spacetime is equivalent to other eight-dimensional descriptions of spacetime, such as the biquaternions\cite{Demir2011} and the octonions\cite{gogberashvili2005octonionic,gogberashvili2006octonionic}.

In order to produce a suitable metric distance we define Clifford conjugation as
\be
\cliffconj{X} = c t - \boldsymbol{x} - \iGAj \boldsymbol{n} + \iGAj c b ,
\ee
which is analogous to the operation of raising and lowering indexes found in four-vector notation.
We can then find a spacetime distance squared $ dS^2 = X \cliffconj{X} $ where
\be \label{FullMetricSpacetimeDistance}
X \cliffconj{X}  = c^2 t^2 - \boldsymbol{x}^2 + \boldsymbol{n}^2 - c^2 b^2 + 2 \iGAj ( c^2 b t - \boldsymbol{x} \cdot \boldsymbol{n} ) .
\ee
We can see the Minkowski invariant interval $ c^2 t^2 - \boldsymbol{x}^2 $ appearing as a special case when the additional spin terms $ \boldsymbol{n} $ and $ b $ are zero.  However, assuming that these terms are indeed present in spacetime then the Minkowski interval is recovered if we select $ b= \pm || \boldsymbol{n} ||/c $ and $ c^2 b t - \boldsymbol{x} \cdot \boldsymbol{n}=0  $, where 
$ || \boldsymbol{n} || =\sqrt{\boldsymbol{n}^2 } = \sqrt{ \boldsymbol{n} \cdot \boldsymbol{n} } = \sqrt{n_1^2 + n_2^2 + n_3^2} $. Substituting the result for $ b $ we find $ \boldsymbol{x}\cdot\boldsymbol{n}= \pm c || \boldsymbol{n} || t $ and dividing through by $ || \boldsymbol{n} || $ we produce
\be
\boldsymbol{x}\cdot \hat{\boldsymbol{n}}= \pm c t ,
\ee
where $ \hat{\boldsymbol{n}} $ is the unit vector in the direction of $ \boldsymbol{n} $. Hence, we have produced a relation describing Einstein's light cone. Now, $ \boldsymbol{x} $ has its minimum value of  $ c t $ when it is parallel to $ \hat{\boldsymbol{n}} $ and so this relation enforces a space-like condition.  In differential form we have $ (d \boldsymbol{x}) \cdot \hat{\boldsymbol{n}}= \pm c d t $ or $ \boldsymbol{v}\cdot \hat{\boldsymbol{n}}= \pm c $. We can see that this relation requires a particle moving at speed $ || \boldsymbol{v} || = c $ with motion parallel to its spin axis $ \hat{\boldsymbol{n}} $. This condition is indeed satisfied for transverse electromagnetic waves as described by Maxwell's equations. Hence, the generalized metric in Eq.~(\ref{FullMetricSpacetimeDistance}) is also null for light, that is, $ d S^2 =  0 $, the same as the conventional Minkowski metric.  Hence, the Minkowski four-dimensional spacetime arises as a special case when it is constructed on the basis of light propagation, in accordance with special relativity and ESC.  

Therefore, in addition to the Lorentz transformations in Eq.~(\ref{spacetimeLorentzTransformations}) we need to add four additional transformations for the four new components of the spacetime events in Eq.~(\ref{multivector3DSimple}) 
\bea  \label{spinLorentzTransformations}
n_1' & = & \gamma (n_1 - v b) \\ \nonumber
n_2' & = & n_2 \\ \nonumber
n_3' & = & n_3 \\ \nonumber
b' & = & \gamma (b - v n_1/c^2 ).  \nonumber
\eea
These transformations follow from the identification of the trivector with helicity and the bivector with spin. That is, for an object initially at rest spinning in a plane with angular velocity $ \omega $ but then boosted to a velocity $ \boldsymbol{v} $, will produce an amount of helical motion $ \boldsymbol{v} \cdot \boldsymbol{w}  $.  Hence, transforming the helicity we find $ h' d t  = \gamma (h - v w/c^2 ) d t $ as shown in Eq.~(\ref{spinLorentzTransformations}) where $ h = d b/d t $ and $ w = dn/dt $.
For a circularly polarized photon with angular frequency $ \boldsymbol{w} $, we have $ \boldsymbol{v} = \boldsymbol{c} $ parallel to the spin axis $ \boldsymbol{w} $ and so we would generate helicity $ h = c w/c^2 = w/c $ radians per meter.  That is, $ d b = d n /c $, producing the result obtained earlier of $ b^2 = \boldsymbol{n}^2/c^2 $, for light, and so producing a null contribution to the metric from the spin terms.  This indicates from the fourth equation in Eq.~(\ref{spinLorentzTransformations}), that photons have no intrinsic helicity $ b $, and so in a hypothetical rest frame we would simply observe a rotating electric field of angular frequency $ \boldsymbol{w} $ with the magnetic arising from the Lorentz boost of the electric field.

We now wish to confirm the invariance of the generalized interval in Eq.~(\ref{FullMetricSpacetimeDistance}) between the two frames with these transformations.
We find firstly
\bea \nonumber
{\boldsymbol{n}'}^{2} - c^2 {b'}^{2} & = & \frac{ (n_1 - v b)^2 + n_2^2 + n_3^2- c^2 (b - v n_1/c^2 )^2 }{1 - v^2/c^2 } \\ 
& = & \boldsymbol{n}^{2} - c^2 b^{2} 
\eea
and for the imaginary components we find
\bea
& & c^2 b' t' - \boldsymbol{x}' \cdot \boldsymbol{n}'  \\ \nonumber
& = & \frac{ c^2 ((b - v n_1/c^2 )(t - v x_1/c^2 ) )- (x_1 - v t)(n_1 - v b) }{1 - v^2/c^2}  \\ \nonumber
& = & c^2 b t - \boldsymbol{x} \cdot \boldsymbol{n} ,
\eea
which are invariant, as required. Therefore, the transformations given in Eq.~(\ref{spacetimeLorentzTransformations}) and Eq.~(\ref{spinLorentzTransformations}) keeps the generalized metric distance shown in Eq.~(\ref{FullMetricSpacetimeDistance}) invariant.

Now, the conventional boost operation, shown in Eq.~(\ref{spacetimeLorentzTransformations}) and Eq.~(\ref{spinLorentzTransformations}) can be written efficiently in GA\cite{Chappell2014IEEE} with the single operation
\be \label{ConventionalBoostGA}
X' = \rme^{-\phi {\hat{\boldsymbol{v}}/2}} X \rme^{-\phi {\hat{\boldsymbol{v}}/2} } ,
\ee
where we define $ \tanh \phi = v/c $ with $ v = \sqrt{\boldsymbol{v}^2} $ as the magnitude of the relative velocity between the frames.  We then find $ \cosh \phi  = \sqrt{1-v^2/c^2} = \gamma $ and $ \sinh \phi = \gamma v/c $, as required.
The full Lorentz group also includes rotations that can be represented as
\be \label{ConventionalRotationsGA}
X' = \rme^{-\theta {\hat{\boldsymbol{w}}/2}} X \rme^{\theta {\hat{\boldsymbol{w}}/2} } ,
\ee
where $ \theta $ is the rotation angle and $ \hat{\boldsymbol{w}} $ is the rotation axis.

Therefore, the conventional Lorentz boost given in Eq.~(\ref{ConventionalBoostGA}), produces the transformations in Eq.~(\ref{spacetimeLorentzTransformations}) and Eq.~(\ref{spinLorentzTransformations}),  and keeps the metric distance in Eq.~(\ref{FullMetricSpacetimeDistance}) invariant.   We will now show that these more general eight-dimensional spacetime events, shown in Eq.~(\ref{multivector3DSimple}), also allow a more general class of Lorentz transformations than the conventional ones shown in Eq.~(\ref{ConventionalBoostGA}) and Eq.~(\ref{ConventionalRotationsGA}).

\section{Generalized transformations}

We define a bilinear transformation on a multivector $ M $ as
\be \label{generalTransformation}
M' = R M S ,
\ee
where $ M',M, R, S \in C\ell(\Re^3) $.
We then find, using Eq.~(\ref{FullMetricSpacetimeDistance}), the transformed multivector amplitude
\be \label{invarianceMultivectorTransformation}
| M' |^2 = R M S \, {\overline{R M S}} = R M S \cliffconj{S} \cliffconj{M} \cliffconj{R} = |R|^2 |S|^2 | M |^2 ,
\ee
where we have used the anti-involution property of Clifford conjugation and the fact that the amplitude is commuting with respect to the algebra. Hence, provided we specify a unitary condition $ |R|^2 |S|^2 = 1 $ for these transformations, then the amplitude $ | M | $ will be invariant. Without loss of generality, this constraint can be satisfied writing the operators $ R $ and $ S $ in exponential form~\cite{Hestenes:1966,Hestenes2003} as
\be \label{HomLorentzGroup}
X' = \rme^{ \boldsymbol{p} + \iGAj \boldsymbol{q}  } X \rme^{ \boldsymbol{r} + \iGAj \boldsymbol{s} },
\ee
where $ \boldsymbol{p}, \boldsymbol{q},  \boldsymbol{r}, \boldsymbol{s} $ are three-vectors.
With $ R = \rme^{ \boldsymbol{p} + \iGAj \boldsymbol{q}  } $ we find $ R \cliffconj{R} = \rme^{ \boldsymbol{p} + \iGAj \boldsymbol{q}  } \rme^{ -\boldsymbol{p} - \iGAj \boldsymbol{q}  } = 1 $, as required.
We thus have an expanded transformation group with twelve free parameters, as compared with the conventional six-dimensional Lorentz group, which now appears as a special case, as shown in Eq.~(\ref{ConventionalBoostGA}) and Eq.~(\ref{ConventionalRotationsGA}).

Hence the Minkowski interval forming the conventional constraint in Eq.~(\ref{TwoObserverLightCones}), when generalized with the interval in Eq.~(\ref{FullMetricSpacetimeDistance}), allows the Lorentz transformations to be generalized to  Eq.~(\ref{HomLorentzGroup}).

\section{Invariance of Maxwell's equations}

Maxwell's electromagnetic field equations are conventionally written\cite{Griffiths:1999} as the four equations
\bea \label{MaxwellClassical}
\mathbf{\nabla} \cdot \mathbf{E} & = & \frac{\rho}{\epsilon},   \,\,\,\,\,\,\, \rm{(Gauss\text{'} \,\, law)} ; \\ \nonumber
\mathbf{\nabla} \times \mathbf{B} - \frac{1}{c^2} \frac{\partial \textbf{E} }{\partial t} & = & \mu_0 \mathbf{J},  \,\,\,  \rm{(Amp\grave{e}re\text{'}s \,\, law)} ; \\ \nonumber
\mathbf{\nabla} \times \mathbf{E} +  \frac{\partial \mathbf{B}}{\partial t}   & = &  \mathbf{0} , \,\,\,\,\,\,\,\,\, \rm{(Faraday\text{'}s \,\, law)} ; \\ \nonumber
\mathbf{\nabla} \cdot \mathbf{B}  & = &  0 , \,\,\,  \,\,\,  \,\,\,  \rm{(Gauss\text{'} \,\, law \,\, of \,\, magnetism)} ,  \nonumber
\eea
where $ \nabla = e_1 \partial_x + e_2 \partial_y + e_3 \partial_z $. 

Using Clifford geometric algebra, Maxwell's four equations can be written as a single equation~\cite{Chappell2014IEEE,Baylis2001}
\be
\left (\frac{1}{c  } \frac{\partial}{\partial t } + \nabla \right) F = \frac{\rho}{\epsilon} - \mu c \mathbf{J} ,
\ee
where the electromagnetic field is $  F = \boldsymbol{E} + \iGAj c \boldsymbol{B}  $.
Now, writing $ \partial = \frac{1}{c} \partial_{t} + \nabla $ and $ J = \frac{\rho}{\epsilon} - \mu c \mathbf{J} $ we can then write
\be \label{MaxwellsEquationsClifford}
\partial F = J.
\ee
Now $ \partial $ and $ J $ are known four-vectors and so we can act with our general transformation in Eq.~(\ref{HomLorentzGroup}), so that $ \partial' = R \partial S $ and $ J' = R J S $ and produce
\be \label{TransformedMaxwell}
R \partial S \cliffconj{S} F S = R J S  ,
\ee
where we have inserted $ \cliffconj{S} $ and $ S $ around $ F $ for self consistency. That is, we know $ R \cliffconj{R} = S \cliffconj{S} = 1 $ and so multiplying Eq.~(\ref{TransformedMaxwell}) from the left by $ \cliffconj{R} $ and from the right by $ \cliffconj{S} $ we return to Maxwell's equations in their form before transformation, as shown in Eq.~(\ref{MaxwellsEquationsClifford}).  As we see this requires a field transformation $ F' = \cliffconj{S} F S $, specifically
\be \label{ExpandedGeneralEMFieldGroup}
\boldsymbol{E}' + \iGAj c \boldsymbol{B}' = \rme^{ -\boldsymbol{r} - \iGAj \boldsymbol{s}  } ( \boldsymbol{E} + \iGAj c \boldsymbol{B} ) \rme^{ \boldsymbol{r} + \iGAj \boldsymbol{s}  } .
\ee
The bivector contribution $ \iGAj \boldsymbol{s} $ produces a rotation of the field and the three-vector component $ \boldsymbol{r} $ produces a boost.
This is, in fact, equivalent to the standard transformation for the electromagnetic field.     

We can put the operators in Eq.~(\ref{HomLorentzGroup})  in a more explicit form by writing $ G = \boldsymbol{p} + \iGAj \boldsymbol{q}  $ so that we find
\be \label{ExplicitExponential}
\rme^{ \boldsymbol{p} + \iGAj \boldsymbol{q}  } = \rme^G = \cos | G | + \hat{G} \sin  | G | ,
\ee
where $ G^2 = -| G|^2 = \boldsymbol{p}^2 -  \boldsymbol{q}^2 + 2 \iGAj \boldsymbol{p} \cdot \boldsymbol{q} $ and $ \hat{G} = G/|G| $. We therefore have 
\be \label{AmplitudeG}
| G| = \sqrt{-\boldsymbol{p}^2 +  \boldsymbol{q}^2 - 2 \iGAj \boldsymbol{p} \cdot \boldsymbol{q}  },
\ee
which will produce a complex-like number in general.  We can see that the trigonometric functions, and indeed the square root, may need to act over the field of complex-like numbers $ a + \iGAj b $, where $ \iGAj $ is the trivector.  However, as $ \iGAj^2 = - 1 $ a commuting scalar we can simply utilize standard expressions from complex number theory. 
Note that $ \hat{G}^2 = -1 $ and so acts like a unit imaginary and so Eq.~(\ref{ExplicitExponential}) is analogous to an Euler-type relation $ \rme^{ \iGAj \theta } = \cos \theta + \iGAj \sin \theta $. 

For example, for pure boosts with $ \boldsymbol{q} = 0  $  we have $ | G| = \iGAj p $, where $ \iGAj $ is the trivector, and so for a boost of a pure electric field we have
\bea
&  & \big (\cos | G | - \hat{G} \sin  | G | \big ) \boldsymbol{E} \big ( \cos | G | + \hat{G} \sin  | G | \big )   \\ \nonumber
& = & \cos^2 | G |  \boldsymbol{E} - \sin^2 | G |  \hat{G} \boldsymbol{E} \hat{G} \\ \nonumber
&  & + \cos | G | \sin  | G | (\boldsymbol{E} \hat{G}- \hat{G} \boldsymbol{E} )  \\ \nonumber
& = & \cosh^2 p \boldsymbol{E} - \sinh^2 p \hat{\boldsymbol{p}} \boldsymbol{E} \hat{\boldsymbol{p}} - \sinh 2 p  \left (\boldsymbol{E} \wedge \hat{\boldsymbol{p}} \right ) ,  \nonumber
\eea
where $ \hat{\boldsymbol{p}}  = \boldsymbol{p}/p $. 
Now with $ \tanh \phi = v/c $ and $ 2 \boldsymbol{p} = \phi \hat{\boldsymbol{v}} $, where $ \hat{\boldsymbol{v}}  = \boldsymbol{v}/v $, then 
\bea
E' & = & \cosh \phi \boldsymbol{E}_{\perp} + \boldsymbol{E}_{\parallel} + \sinh \phi  \left (\boldsymbol{E} \wedge \boldsymbol{v} \right )  \\ \nonumber
& = & \gamma \boldsymbol{E}_{\perp} + \boldsymbol{E}_{\parallel} + \gamma  \boldsymbol{E} \wedge \boldsymbol{v}/c ,  \nonumber
\eea
using $ \gamma = \cosh \phi $.
We can see a magnetic field arising from  $ \boldsymbol{E} \wedge \boldsymbol{v}/c = - \iGAj \boldsymbol{E} \times \boldsymbol{v}/c $.  Now, for the field variable $ \boldsymbol{E}' + \iGAj c \boldsymbol{B}' $ we therefore have $ \boldsymbol{B}' = \iGAj \boldsymbol{E} \times \boldsymbol{v}/c^2 $, agreeing with the conventional transformation of the electric field.  We also find the perpendicular field components increased by $ \gamma $, as expected.

Hence the more general coordinate transformation in Eq.~(\ref{HomLorentzGroup}) nevertheless still requires the conventional field transformation, shown in Eq.~(\ref{ExpandedGeneralEMFieldGroup}).

\section{Transforming the sources}

We have found that Maxwell's equations are invariant under the generalized Lorentz transformations in Eq.~(\ref{HomLorentzGroup}), and the transformation of the field remains the conventional one.  We now consider the effect of the generalized transformations on the four-current sources.
Maxwell's equations, with the presence of magnetic monopoles become\cite{Dirac1948}
\be
\left (\frac{1}{c  } \frac{\partial}{\partial t } + \nabla \right) F = \frac{\rho}{\epsilon} - \mu c \boldsymbol{J} -  \iGAj \mu_0 \boldsymbol{J^m} + \iGAj c \mu_0 \rho^m,
\ee
where $ \rho^m $ and $ \boldsymbol{J^m} $ are the monopole sources and currents, respectively.
With this equation Maxwell's equations will now be modified to $ \nabla \cdot \boldsymbol{B} =  \rho^m $ and $ \boldsymbol{\nabla} \times \boldsymbol{E} +  \partial_t \boldsymbol{B} = - \boldsymbol{J}^m $, in agreement with conventional results.

Now, for the four-current  $ J = \rho/\epsilon - \mu c \mathbf{J} $ then the general transformation will produce in general
\be
R \left ( \rho/\epsilon  -\mu c \boldsymbol{J} \right ) S = \rho'/\epsilon -\mu c \boldsymbol{J}' + \iGAj \boldsymbol{K}' + \iGAj \kappa' ,
\ee
where $ \iGAj \kappa' $ is a magnetic monopole source and $ \iGAj \boldsymbol{K}' = \iGAj (k_1 e_1 + k_2 e_2 + k_3 e_3 ) $ is a monopole  current. 

Maxwell's equation assumes that magnetic monopole sources are identically zero and indeed, despite extensive experimental searches magnetic monopoles have never been conclusively observed~\cite{Milton2006}.  Hence, if we follow Maxwell and accept the non-existence of monopoles then this implies that the generalized transformations need to be restricted by this condition to the conventional ones.  
Note also that the extension of Maxwell's equations to describe massive photons will also require the generalized transformations to be restricted to the conventional ones as it relies on the four potential $ A = \phi - c \boldsymbol{A} $, which transforms analogous to the four current. Incidentally, Maxwell's equations can be written in this case as
$ \partial F = J - m^2 A $, where $ m $ is the presumed mass of the photon.

We note, that while the full generalized Lorentz transformations may be ruled out on these physical grounds, perhaps a limited extension of the conventional Lorentz transformations is feasible, which will be explored next.

\subsection{A limited generalization}

For a generalized Lorentz transformation to be consistent with Maxwell's equations then we need to incorporate the non-existence of monopole sources.  This implies that a boost of a current source must leave the bivector and trivector terms zero.

We firstly define the involution of {\it reversion}, that reverses the order of all products and produces
\be
\tilde{X} = c t + \boldsymbol{x} - \iGAj \boldsymbol{n} - \iGAj c b .
\ee

Hence, if the bivector and trivector terms are absent we have
\be
X = \tilde{X} .
\ee

Now, for a general boost, we find
\be
J' = \rme^{ \boldsymbol{p} + \iGAj \boldsymbol{q}  }  J \rme^{ \boldsymbol{r} + \iGAj \boldsymbol{s} } .
\ee
Therefore, if we require an absence of bivector and trivector sources, we require $ J' = \tilde{J'} $ or
\be
\rme^{ \boldsymbol{p} + \iGAj \boldsymbol{q}  } J \rme^{ \boldsymbol{r} + \iGAj \boldsymbol{s} } = \rme^{ \boldsymbol{r} - \iGAj \boldsymbol{s} } J \rme^{ \boldsymbol{p} - \iGAj \boldsymbol{q}  } ,
\ee
where we have used the fact that $ J = \tilde{J} $. 
As this must be true for an arbitrary current $ J $, we therefore require $ \boldsymbol{r} = \boldsymbol{p} $ and $ \boldsymbol{s} = - \boldsymbol{q} $.
This then produces the operator
\be  \label{GeneralLorentzNoMonopoles}
J' = \rme^{\boldsymbol{p} - \iGAj \boldsymbol{q}  } J \rme^{\boldsymbol{p} + \iGAj \boldsymbol{q} } ,
\ee
which is the most general transformation that ensures monopoles sources remain zero.
Note the similarity but also the single sign difference compared with the field operators in Eq.~(\ref{ExpandedGeneralEMFieldGroup}).

Note that the generalized boost operation in Eq.~(\ref{GeneralLorentzNoMonopoles}) is in fact still standard physics, and has the physical application of the Thomas rotation~\cite{TaylorWheeler,chappell2011revisiting}.  This phenomena arises from the fact that two non-parallel boosts induces an apparent rotation.  The bivector terms in Eq.~(\ref{GeneralLorentzNoMonopoles}), being rotation operators, thus describe this aspect of the Lorentz boosts.

For this combined boost operator we have
\be
\rme^{ \phi  (\boldsymbol{p} \pm \iGAj \boldsymbol{q} )/\sqrt{(\boldsymbol{p} \pm \iGAj \boldsymbol{q})^2 }  } = \cosh \phi +  \frac{\boldsymbol{p} \pm \iGAj \boldsymbol{q} }{\sqrt{(\boldsymbol{p} \pm \iGAj \boldsymbol{q})^2 } } \sinh \phi
\ee
where  $ \tanh \phi = \sqrt{(\boldsymbol{p} + \iGAj \boldsymbol{q})^2 } $, which upon rearrangement implies $ \cosh \phi = 1/\sqrt{1 - \boldsymbol{p}^2 + \boldsymbol{q}^2 - 2 \iGAj  \boldsymbol{p} \cdot  \boldsymbol{q} } $.  For the special case where  $ \boldsymbol{q} = 0 $, we revert to the standard boost operation, shown in Eq.~(\ref{ConventionalBoostGA}).

\subsubsection{Effect on time}

We now use transformations that are compliant with Maxwell's equations, shown in Eq.~(\ref{GeneralLorentzNoMonopoles}), to find the effect on time within a coordinate transformation.
Transforming from the rest frame we find for the scalar part represented by $ \langle \cdots \rangle_0 $, which gives the time coordinate in the new frame
\bea \nonumber
& = & \Big \langle \big (\cos | G| + \hat{G} \sin  | G | \big ) \tau \big ( \cos | H | + \hat{H} \sin  | H | \big ) \Big \rangle_0 \\ \nonumber
& = & \tau \left ( \cos | G| \cos | H | +   \frac{\sin | G| \sin | H |}{|G||H| } \left \langle G H \right \rangle_0 \right ), 
\eea
where $ \tau $ is the time in the rest frame.
Now $ G H = (\boldsymbol{p}+\iGAj \boldsymbol{q})(\boldsymbol{p}-\iGAj \boldsymbol{q}) = \boldsymbol{p}^2+\boldsymbol{q}^2 - 2 \iGAj ( \boldsymbol{p} \wedge \boldsymbol{q} ) $ and from Eq.~(\ref{AmplitudeG}) we can write 
\be \label{magGalpha}
| G | = \alpha_+ + {\rm{sgn}} (\boldsymbol{p} \cdot \boldsymbol{q}) \iGAj \alpha_- ,
\ee
where 
\be \label{AlphaPlusMinus}
\alpha_{\pm} =  \pm \frac{1}{\sqrt{2}}\sqrt{\sqrt{(\boldsymbol{q}^2-\boldsymbol{p}^2)^2 +  4 (\boldsymbol{p} \cdot \boldsymbol{q})^2 } \pm (\boldsymbol{q}^2-\boldsymbol{p}^2 ) } 
\ee
are real scalars and the sign of the imaginary term, in Eq.~(\ref{magGalpha}), is given by the sign of $ \boldsymbol{p} \cdot \boldsymbol{q} $.  We therefore have $ | H | = \alpha_+ - {\rm{sgn}} (\boldsymbol{p} \cdot \boldsymbol{q}) \iGAj \alpha_- $.  
Now, using standard trigonometric identities, shown in Appendix~B, we find
\bea
t & = & \tau (\cos^2 \alpha_+ \cosh^2 \alpha_- + \sin^2 \alpha_+ \sinh^2 \alpha_-) \\ \nonumber
& & + \tau (\sin^2 \alpha_+ \cosh^2 \alpha_- + \cos^2 \alpha_+ \sinh^2 \alpha_-) \frac{\boldsymbol{p}^2+\boldsymbol{q}^2}{\alpha_+^2 + \alpha_-^2}  .  \nonumber
\eea
We can also find $ \alpha_+^2 + \alpha_-^2 = +\sqrt{(\boldsymbol{q}^2-\boldsymbol{p}^2)^2 + 4 (\boldsymbol{p} \cdot \boldsymbol{q})^2 } $. 

To confirm a correspondence with regular boosts we find with $ \boldsymbol{q} = 0 $ that  $ \alpha_+ = 0 $ and $ \alpha_- = p $.  Therefore
\be
t = \tau (\cosh^2 p + \sinh^2 p) = \tau \cosh 2 p .
\ee
For $ \tanh 2 p = v/c $ then $ \cosh 2 p = 1/\sqrt{1 - v^2/c^2 } $, which is the conventional $ \gamma $ factor for time dilation.

If we write $ t = \gamma_g \tau $ for the generalized time dilation then we have
\bea
\gamma_g & = & \cos^2 \alpha_+ \cosh^2 \alpha_- + \sin^2 \alpha_+ \sinh^2 \alpha_- \\ \nonumber
& & +  \left (\sin^2 \alpha_+ \cosh^2 \alpha_- + \cos^2 \alpha_+ \sinh^2 \alpha_- \right ) \frac{\boldsymbol{p}^2+\boldsymbol{q}^2}{\alpha_+^2 + \alpha_-^2} . \nonumber
\eea
We can show that $ 1 \le \gamma_g \le \gamma $ and so the generalized time dilation is bounded by the conventional values.  That is, the inclusion of rotational terms into the boost operation tends to reduce the amount of time dilation.

\section{Thomas rotation}

For the case of two consecutive boosts we have the operators
\be \label{ThomasRotation}
R = \rme^{ - \phi_2 \hat{\mathbf{v}}_2/2}  \rme^{ - \phi_1 \hat{\mathbf{v}}_1/2 } = \rme^{ -  \phi_c \hat{\mathbf{v}}_c/2 } \rme^{  - \iota \theta/2 } ,
\ee
where we have combined the two boosts\cite{chappell2011revisiting} into a single boost $ \phi_c \hat{\mathbf{v}}_c $ and a rotation $ \theta $, where
\be
\tan \frac{\theta}{2} = \frac{ \sin \delta \sinh \frac{\phi_1}{2} \sinh \frac{\phi_2}{2} }{\cos \delta \sinh \frac{\phi_1}{2} \sinh \frac{\phi_2}{2} - \cosh \frac{\phi_1}{2} \cosh \frac{\phi_2}{2} } ,
\ee 
where $ \delta $ is the angle between the boost directions, given by $ \cos \delta = \hat{\mathbf{v}}_1 \cdot \hat{\mathbf{v}}_2 $. Hence we can see that only for parallel boosts, that is $ \delta = 0 $, will there not in fact be a Thomas rotation $ \theta $, of the frame.  We also have $ \iota = \hat{\mathbf{v}}_2 \wedge \hat{\mathbf{v}}_1/ \sin \delta $ being the unit bivector in the plane of the two boosts.
We thus see that for two non-parallel boosts there is an implied rotation in the plane of the two boosts.  However, we note that our boost transformation in Eq.~(\ref{GeneralLorentzNoMonopoles}) that as $ \boldsymbol{p} $ and $ \boldsymbol{q} $ are not necessarily coplanar then this allows a rotation out of the plane.
This case would therefore require three non-coplanar boosts that would then create an implied helical motion of the particle.  That is, we have
\be \label{ThomasRotation3D}
R = \rme^{ - \phi_3 \hat{\mathbf{v}}_3/2} \rme^{ - \phi_2 \hat{\mathbf{v}}_2/2}  \rme^{ - \phi_1 \hat{\mathbf{v}}_1/2 } = \rme^{ -  \phi_c \hat{\mathbf{v}}_c/2 } \rme^{  - \iota \theta/2 } \rme^{  - \iota^{\perp} \psi/2 },
\ee
where we now have two rotations, one in the plane of the combined boost and an additional one $ \iota^{\perp} $ creating a spin vector parallel to the combined boost direction forming helical motion.  Note, though, that while we can neatly describe the Thomas rotation in the plane, as well as a generalization  in three-dimensions, nevertheless we remain within conventional physics.

\subsection{The Dirac equation}

The Dirac relativistic wave equation for the electron is commonly written
\be
\gamma^{\mu} \partial_{\mu }  \psi =  -i m \psi ,
\ee
where $ \gamma^{\mu} $, for $ \mu \in \{0,1,2,3\} $, are the 
four anti-commuting gamma matrices with $ \gamma^{\mu} \gamma^{\nu} +  \gamma^{\nu} \gamma^{\mu} =  2 g_{\mu \nu } I $, where $ g_{\mu \nu } = {\rm{diag}}[-1,1,1,1] $ is the Minkowski metric.  Note that we have now chosen units in which $ c = \hbar = 1 $. The similarity between the anti-commuting Dirac basis matrices $ \gamma^{\mu} $ and the anti-commuting basis vectors defined in $ C\ell \left(\Re^3 \right) $ allows us to write an isomorphic equation in GA
\be \label{DiracGA}
\left (\partial_t + \nabla \right ) \psi  = -m \psi^{*} \iGAj e_3 ,
\ee
where $ \psi \in C\ell \left(\Re^3 \right) $.  Note that with a wave function multivector $ \psi = a + \boldsymbol{E} + \iGAj \boldsymbol{B} + \iGAj b $ then we define the involution $ \psi^{*} = a - \boldsymbol{E} + \iGAj \boldsymbol{B} - \iGAj b $.
We note that the space of multivectors in $ C\ell \left(\Re^3 \right) $ is eight dimensional and so can be made isomorphic to the eight-dimensional Dirac spinor, as required.  
We note that the left hand side of the Dirac and Maxwell equations are identical and so we require the same transformation $ R \partial S \cliffconj{S} \psi \cliffconj{T} $, where we define $ T =\rme^{ \boldsymbol{v} + \iGAj \boldsymbol{w}  } $. Therefore $ \psi' = \cliffconj{S} \psi \cliffconj{T} $ and so
\be
\psi^{*'} =  \cliffconj{S}^* \psi^* \cliffconj{T}^* .
\ee
So writing the transformation out in full, we find
\be
\rme^{ \boldsymbol{p} + \iGAj \boldsymbol{q}  } \partial \psi \rme^{ -\boldsymbol{v} - \iGAj \boldsymbol{w}  } = - m \rme^{ \boldsymbol{r} - \iGAj \boldsymbol{s}  } \psi^* \rme^{ \boldsymbol{v} - \iGAj \boldsymbol{w}  } \iGAj e_3.
\ee
In order to recover the untransformed Dirac equation we therefore require $ \rme^{ \boldsymbol{p} + \iGAj \boldsymbol{q}  } = \rme^{ \boldsymbol{r} - \iGAj \boldsymbol{s}  } $ or $ \boldsymbol{r} = \boldsymbol{p} $ and  $ \boldsymbol{q} = -\boldsymbol{s} $, which is the standard Lorentz boost, shown previously in Eq.~(\ref{GeneralLorentzNoMonopoles}).  Hence, the Dirac equation also enforces the standard Lorentz boost over four-vectors, although in this case apparently unrelated to the presence of monopoles as it was for Maxwell's equations.

\section{Conclusion}

We find a generalized Lorentz transformation in Eq.~(\ref{HomLorentzGroup}) that preserves the Minkowski invariant interval as well as retaining the form of Maxwell's equations including the expected field transformation.  We then find that in order to incorporate the absence of magnetic monopoles, as required by Maxwell's equations, we need to restrict the transformations to Eq.~(\ref{GeneralLorentzNoMonopoles}).  This requirement then enforces the conventional Lorentz transformations.  We also then find that the Dirac equation also enforces the conventional Lorentz boost, although in this case unrelated to the existence of monopoles.
Nevertheless, while the conventional Lorentz boosts are still required, they can now act over a generalized eight-dimensional spacetime that now includes the rotational degrees of freedom.

We can now trace how the Minkowski spacetime and Lorentz transformations have arisen.  Firstly, we note that Maxwell's equations imply the Lorentz transformation as well as transverse electromagnetic waves, which then produces the space and time transformations of special relativity based on light signaling and finally summarized in the four-dimensional spacetime continuum of Minkowski.  However, we can see that the Minkowski four-dimensional spacetime only arises from the more general eight-dimensional structure we shown in Eq.~(\ref{multivector3DSimple}) when light is utilized to establish space and time coordinates.  That is, the bivector and trivector spin contributions to the metric, in Eq.(\ref{GeneralLorentzNoMonopoles}), are identically zero in this case, thus reducing to Minkowski spacetime.  However, for a full description of non-lightlike massive particles, where the additional spin components are not identically zero, then the full eight-dimensional description shown in Eq.~(\ref{multivector3DSimple}) will be required.
Indeed there are several arenas of physics where the standard spacetime coordinates seem to fail, such as inside the event horizon of black holes or the inability to establish local realism in quantum mechanics.  Also, the photon is the only lightlike particle known within the standard model that can be used to establish a spacetime framework, except for perhaps the graviton or the gluon. 

We note that while the generalized transformations appeared ruled out by the non-existence of magnetic monopoles, there are still several lines of inquiry that can be pursued.  Firstly, while true magnetic monopoles may not exist in nature, quasi magnetic monopoles have been detected in spin-ice systems\cite{Kadowaki2009}.  Secondly, the generalized spacetime events may also produce additional interaction terms that then appear in the Lorentz force law.   Other possible applications are investigations of the longitudinal electrodynamic force and the Abraham-Minkowski controversy regarding the correct definition for light momenta in dielectrics, as this issue is related to the conservation of angular momentum\cite{johansson1996longitudinal,obukhov2008electromagnetic}.
Our final coordinate transformation in Eq.~(\ref{GeneralLorentzNoMonopoles}) encapsulates the properties of the Thomas rotation in a single operation that consists of two non-parallel boosts.  We also find that our operator is able to describe a more general Thomas rotation involving three non-coplanar boosts with an implied helical motion.

Finally, while Maxwell's electrodynamic equations force us to restrict the use of the general transformation to the conventional one, we know that Maxwell's equations formed a classical theory before quantization and before the advent of general relativity. Hence our generalized structure and transformations may have applicability to describe quantum effects or other fundamental forces and provide a framework to explore the theoretical aspects of magnetic monopoles.  If monopoles are indeed found to exist in nature then the generalized transformations that we describe are applicable and so would allow a much greater range of space, time and field transformations than are currently permitted by the Lorentz group.

\section{Acknowledgment}

N. Iannella was supported by the People Programme (Marie Curie Actions) of the European Unions Seventh Framework Programme (FP7/2007-2013) under REA grant agreement No PCOFUND-GA-2012-600181.

\appendix

\section{The multivector products}

In Clifford geometric algebra we form the space of multivectors $ \Re \oplus \Re^3 \oplus \bigwedge^2 \Re^3 \oplus \bigwedge^3 \Re^3 $, an eight-dimensional real vector space denoted by $ C\ell(\Re^3) $.  This thus consists of the sum of a scalar, vector, bivector and trivector.
Defining vectors $ \boldsymbol{v} = v_1 e_1 + v_2 e_2 + v_3 e_3 $ and $ \boldsymbol{u} = u_1 e_1 + u_2 e_2 + u_3 e_3 $, where $ v_i, u_i \in \Re $, we find their algebraic product using the distributive law of multiplication over addition as
\bea \label{VectorProductExpand}
\textbf{u} \textbf{v} & = & (e_1 u_1 + e_2 u_2 + e_3 u_3 ) ( e_1 v_1 + e_2 v_2 + e_3 v_3 ) \\ \nonumber
& = & u_1 v_1 + u_2 v_2 + u_3 v_3  + (u_2 v_3 - v_2 u_3 ) e_2 e_3 \\ \nonumber
& & + (u_1 v_3 - u_3 v_1 ) e_1 e_3 + (u_1 v_2 - v_1 u_2 ) e_1 e_2 \\ \nonumber
& = & \textbf{u} \cdot \textbf{v}  + \textbf{u} \wedge \textbf{v}, \nonumber
\eea
which produces a sum of symmetric and antisymmetric products, being the sum of a scalar and a bivector.
We can then write   
\be
\textbf{u} \cdot \textbf{v} = \tfrac{1}{2} (  \textbf{u} \textbf{v} +  \textbf{v} \textbf{u} ) \, , \,\,\, \textbf{u} \wedge \textbf{v} = \tfrac{1}{2} (  \textbf{u} \textbf{v} -  \textbf{v} \textbf{u} ) .
\ee
Also, we find 
\be
\textbf{u} \wedge \textbf{v} = \iGAj \textbf{u} \times \textbf{v} ,
\ee
which forms a connection with the conventional cross product.

\section{Useful trigonometric relations}

We have $ | G | = \alpha_+ + \iGAj \alpha_- $ and similarly $ | H | = \beta_+ + \iGAj \beta_- $. 
Now, using the trigonometric identities $ \cos (a + \iGAj b ) = \cos a \cosh b - \iGAj \sin a \sinh b $ and $ \sin (a + \iGAj b ) = \sin a \cosh b + \iGAj \cos a \sinh b $, we therefore have the following results:
\bea \nonumber
\cos | G | \cos | H | &= & (\cos \alpha_+ \cosh \alpha_- - \iGAj \sin \alpha_+ \sinh \alpha_- ) \\ \nonumber
& & \times (\cos \beta_+ \cosh \beta_- - \iGAj \sin \beta_+ \sinh \beta_- ) \\ 
 & = & \cos \alpha_+ \cosh \alpha_- \cos \beta_+ \cosh \beta_- \\ \nonumber
& & - \sin \alpha_+ \sinh \alpha_- \sin \beta_+ \sinh \beta_- \\ \nonumber
& & - \iGAj (\cos \beta_+ \cosh \beta_- \sin \alpha_+ \sinh \alpha_- \\ \nonumber
& & + \cos \alpha_+ \cosh \alpha_- \sin \beta_+ \sinh \beta_- ) , \nonumber
\eea
\bea \nonumber
 \sin | G | \sin | H | & = & (\sin \alpha_+ \cosh \alpha_- + \iGAj \cos \alpha_+ \sinh \alpha_- ) \\ \nonumber
& & \times (\sin \beta_+ \cosh \beta_- + \iGAj \cos \beta_+ \sinh \beta_- ) \\ 
 & = & \sin \alpha_+ \cosh \alpha_- \sin \beta_+ \cosh \beta_- \\ \nonumber
& & - \cos \alpha_+ \sinh \alpha_- \cos \beta_+ \sinh \beta_- \\ \nonumber
& & + \iGAj (\sin \beta_+ \cosh \beta_- \cos \alpha_+ \sinh \alpha_- \\ \nonumber
& & + \sin \alpha_+ \cosh \alpha_- \cos \beta_+ \sinh \beta_- ) , \nonumber
\eea
and
\bea
& & \sin | G | \cos | H | \\ \nonumber
& = & (\sin \alpha_+ \cosh \alpha_- + \iGAj \cos \alpha_+ \sinh \alpha_- ) \\ \nonumber
& & \times (\cos \beta_+ \cosh \beta_- - \iGAj \sin \beta_+ \sinh \beta_- ) \\ \nonumber
 & = & \sin \alpha_+ \cosh \alpha_- \cos \beta_+ \cosh \beta_- \\ \nonumber
& & + \cos \alpha_+ \sinh \alpha_- \sin \beta_+ \sinh \beta_- \\ \nonumber
& & + \iGAj (\cos \beta_+ \cosh \beta_- \cos \alpha_+ \sinh \alpha_- \\ \nonumber
& & - \sin \alpha_+ \cosh \alpha_- \sin \beta_+ \sinh \beta_- ) . \nonumber
\eea

\section{Example of a generalized boost}

An example of a generalized boost would be
\be \label{TransformationSpecialExample}
X' = \rme^{ \boldsymbol{p} } X \rme^{ \boldsymbol{r} } = \rme^{ -\phi \hat{\boldsymbol{v}} } X \rme^{ -\phi \hat{\boldsymbol{w}} } .
\ee
This is an extension of the conventional boost in which $  \hat{\boldsymbol{w}} =  \hat{\boldsymbol{v}} $, shown in Eq.~(\ref{ConventionalBoostGA}).
For the boost shown we have $ | G | =  | H | = \iGAj \phi $.  Boosting a spacetime vector $ t + \boldsymbol{x} $, where we now take $ c = 1 $, we find from Eq.~(\ref{ExplicitExponential})
\bea
& & (\cosh \phi + \hat{\boldsymbol{v}} \sinh \phi)( t + \boldsymbol{x} ) (\cosh \phi + \hat{\boldsymbol{w}} \sinh \phi)  \\ \nonumber
& = & t ( \cosh^2 \phi + \hat{\boldsymbol{v}} \hat{\boldsymbol{w}} \sinh^2 \phi \\ \nonumber
& & + \hat{\boldsymbol{v}} \sinh \phi \cosh \phi + \hat{\boldsymbol{w}} \cosh \phi \sinh \phi)  \\ \nonumber
& & + \boldsymbol{x}^{\|} ( \cosh^2 \phi + \hat{\boldsymbol{v}} \hat{\boldsymbol{w}} \sinh^2 \phi \\ \nonumber
& & + \hat{\boldsymbol{v}} \sinh \phi \cosh \phi + \hat{\boldsymbol{w}} \cosh \phi \sinh \phi)  \\ \nonumber
& & + \boldsymbol{x}^{\perp} ( \cosh^2 \phi - \hat{\boldsymbol{v}} \hat{\boldsymbol{w}} \sinh^2 \phi\\ \nonumber
& & -\hat{\boldsymbol{v}} \sinh \phi \cosh \phi + \hat{\boldsymbol{w}} \cosh \phi \sinh \phi ) . \nonumber
\eea
For the vector components we find
\bea
\boldsymbol{x}' & = & t ( \hat{\boldsymbol{v}} \sinh \phi \cosh \phi + \hat{\boldsymbol{w}} \cosh \phi \sinh \phi ) \\ \nonumber
& & + \boldsymbol{x}^{\|} ( \cosh^2 \phi + \hat{\boldsymbol{v}} \hat{\boldsymbol{w}} \sinh^2 \phi)  \\ \nonumber
& & + \boldsymbol{x}^{\perp} ( \cosh^2 \phi - \hat{\boldsymbol{v}} \hat{\boldsymbol{w}} \sinh^2 \phi) ,  \nonumber
\eea
which can be re-arranged to
\bea
\boldsymbol{x}' & = & \boldsymbol{x}^{\|} ( \cosh 2 \phi + (\hat{\boldsymbol{v}} \hat{\boldsymbol{w}}-1) \sinh^2 \phi)  \\ \nonumber
& & + \frac{t}{2} ( \hat{\boldsymbol{v}} \sinh 2 \phi + \hat{\boldsymbol{w}} \sinh 2 \phi ) \\ \nonumber
& & + \boldsymbol{x}^{\perp} ( 1 - (\hat{\boldsymbol{v}} \hat{\boldsymbol{w}}-1) \sinh^2 \phi) . \nonumber
\eea
Now, as $ \tanh 2 \phi = v $ we have $ \cosh 2 \phi = \gamma = 1/\sqrt{1-v^2} $ and $ \sinh 2\phi = \gamma v $ and so we find that
\be
\boldsymbol{x}' = \gamma \left ( \boldsymbol{x}^{\|} + \frac{( \boldsymbol{v} + \boldsymbol{w}) }{2} t \right ) +\boldsymbol{x}^{\perp} + \hat{\boldsymbol{v}} \boldsymbol{x} ( \hat{\boldsymbol{w}}-\hat{\boldsymbol{v}} ) \frac{\gamma - 1}{2}  .
\ee
We note that for conventional boosts $ \hat{\boldsymbol{v}} = \hat{\boldsymbol{w}} $ and so we produce the result shown in Eq.~(\ref{spacetimeLorentzTransformations}).  The last term also vanishes in the non-relativistic limit as $ \gamma \rightarrow 1 $.

For the scalar components we find
\bea
& & t ( \cosh^2 \phi + \hat{\boldsymbol{v}} \cdot \hat{\boldsymbol{w}} \sinh^2 \phi) \\ \nonumber
& & + x^{\|} \sinh \phi \cosh \phi + \boldsymbol{x}^{\|} \cdot \hat{\boldsymbol{w}} \cosh \phi \sinh \phi \\ \nonumber
& & +  \boldsymbol{x}^{\perp} \cdot \hat{\boldsymbol{w}} \cosh \phi \sinh \phi ,  \nonumber
\eea
which can be re-arranged to
\bea
t' & = & t ( \cosh 2 \phi + (\hat{\boldsymbol{v}} \cdot \hat{\boldsymbol{w}} - 1) \sinh^2 \phi) \\ \nonumber
& & + \frac{1}{2} x^{\|} \sinh 2 \phi  + \frac{1}{2} \boldsymbol{x} \cdot \hat{\boldsymbol{w}} \sinh 2 \phi \nonumber
\eea
and so we have
\be
t' = \gamma ( t + \frac{1}{2} v ( x^{\|}   + \boldsymbol{x} \cdot \hat{\boldsymbol{w}} ) )  + t (\hat{\boldsymbol{v}} \cdot \hat{\boldsymbol{w}} - 1) \frac{(\gamma - 1)}{2} . 
\ee
Once again with $ \hat{\boldsymbol{w}} = \hat{\boldsymbol{v}} $ the last term is zero and we return the conventional result shown in Eq.~(\ref{spacetimeLorentzTransformations}).

For this generalized boost, we produce the trivector components
\bea
& & \boldsymbol{x}^{\|} \wedge \hat{\boldsymbol{v}} \wedge \hat{\boldsymbol{w}} \sinh^2 \phi - \boldsymbol{x}^{\perp} \wedge \hat{\boldsymbol{v}} \wedge \hat{\boldsymbol{w}} \sinh^2 \phi \\ \nonumber
& = & -\boldsymbol{x} \wedge \hat{\boldsymbol{v}} \wedge \hat{\boldsymbol{w}}  \frac{(\gamma - 1)}{2} . \nonumber
\eea

Therefore, boosting the four current $ \rho + \boldsymbol{J} $ we would produce the trivector term
\be
-\boldsymbol{J} \wedge \hat{\boldsymbol{v}} \wedge \hat{\boldsymbol{w}}  \frac{(\gamma - 1)}{2} . 
\ee
Now, as the trivector represents magnetic monopole charge, then if we require this to be zero we need $ \hat{\boldsymbol{v}} \wedge \hat{\boldsymbol{w}} = 0 $.  This implies $ \hat{\boldsymbol{v}} $ is parallel to $ \hat{\boldsymbol{w}} $ and so we essentially require the conventional boost in Eq.~(\ref{ConventionalBoostGA}).
However in the non-relativistic limit we have $ \gamma \rightarrow 1 $ and so this term also goes to zero.  Hence monopoles will only appear for relativistic boosts in this case.

We also have the bivector components
\be
t \frac{(\gamma-1)}{2} \hat{\boldsymbol{v}} \wedge \hat{\boldsymbol{w}} + \frac{1}{2} \gamma \boldsymbol{J} \wedge (\boldsymbol{w} -\boldsymbol{v} ) .
\ee
This term, representing the monopole current is non-zero even in the non-relativistic limit, though zero for $ \boldsymbol{v} = \boldsymbol{w} $.

\section{Lorentz transformation of the Dirac spinor}

Multiplying from the left by $ \rme^{ -\boldsymbol{p} - \iGAj \boldsymbol{q}  } $ and from the right by $  \rme^{ \boldsymbol{v} + \iGAj \boldsymbol{w}  } $, Dirac's equation is recovered provided
\be \label{e3condition}
\rme^{ \boldsymbol{v} - \iGAj \boldsymbol{w}  } e_3  \rme^{ \boldsymbol{v} + \iGAj \boldsymbol{w}  } = e_3 .
\ee

Now, in quantum electrodynamics we assume that we can produce the four-current $ J = \psi \tilde{\psi} = \rho + \boldsymbol{J} $. If we transform the field, we obtain the transformed current
\be
J' = \rme^{ -\boldsymbol{r} - \iGAj \boldsymbol{s}  } \psi \rme^{ -\boldsymbol{v} - \iGAj \boldsymbol{w}  }  \rme^{ -\boldsymbol{v} + \iGAj \boldsymbol{w}  } \tilde{\psi} \rme^{ -\boldsymbol{r} + \iGAj \boldsymbol{s}  }  . 
\ee
We have thus recovered the correct Lorentz boost of a four-current provided
\be
\rme^{ -\boldsymbol{v} - \iGAj \boldsymbol{w}  }  \rme^{ -\boldsymbol{v} + \iGAj \boldsymbol{w}  } = 1 ,
\ee
or $ \boldsymbol{v}  = 0 $.  Using this result in Eq.~(\ref{e3condition}) we find the condition
\be \label{e3conditionSimplified}
\rme^{ - \iGAj \boldsymbol{w}  } e_3  \rme^{ \iGAj \boldsymbol{w}  } = e_3 ,
\ee
or $ \boldsymbol{w} = w_3 e_3 $.  Hence, the transformation of the wave function requires the operation $ \psi' = \rme^{ -\boldsymbol{r} - \iGAj \boldsymbol{s}} \psi \rme^{ -\iGAj e_3 w_3  } $ or
\be
\psi' = \rme^{ -\boldsymbol{p} + \iGAj \boldsymbol{q}  } \psi \rme^{ - \iGAj w_3 e_3  } .
\ee

\bibliography{quantum}

\end{document}